\documentclass[conference]{IEEEtran}
\IEEEoverridecommandlockouts
\usepackage{url}
\usepackage{xurl}
\usepackage{cite}
\usepackage{amsmath,amssymb,amsfonts}
\usepackage{algorithmic}
\usepackage{graphicx}
\usepackage{textcomp}
\usepackage{xcolor}
\usepackage{hyperref}
\usepackage{stfloats}
\usepackage{booktabs}

\def\BibTeX{{\rm B\kern-.05em{\sc i\kern-.025em b}\kern-.08em
  T\kern-.1667em\lower.7ex\hbox{E}\kern-.125emX}}
\begin{document}

\title{Physics-Informed Neural Engine Sound Modeling with Differentiable Pulse-Train Synthesis%
\thanks{
Code, model weights, audio: \url{https://rdoerfler.github.io/ptr-model-page/}
}
}

\author{\IEEEauthorblockN{Robin Doerfler}
\IEEEauthorblockA{\textit{Audio Research \& Development} \\
\textit{Impulse Audio Lab GmbH}\\
Munich, Germany \\
robin.doerfler@impulse-audio-lab.com}
\and
\IEEEauthorblockN{Lonce Wyse}
\IEEEauthorblockA{\textit{Music Technology Group} \\
\textit{Universitat Pompeu Fabra}\\
Barcelona, Spain \\
lonce.wyse@upf.edu}
}

\maketitle

\begin{abstract}
Engine sounds originate from sequential exhaust pressure pulses rather than sustained harmonic oscillations. While neural synthesis methods typically aim to approximate the resulting spectral characteristics, we propose directly modeling the underlying pulse shapes and temporal structure. We present the Pulse-Train-Resonator (PTR) model, a differentiable synthesis architecture that generates engine audio as parameterized pulse trains aligned to engine firing patterns and propagates them through recursive Karplus-Strong resonators simulating exhaust acoustics. The architecture integrates physics-informed inductive biases including harmonic decay, thermodynamic pitch modulation, valve-dynamics envelopes, exhaust system resonances and derived engine operating modes such as throttle operation and Deceleration Fuel Cutoff (DFCO). Validated on three diverse engine types totaling 7.5 hours of audio, PTR achieves a 21\% improvement in harmonic reconstruction and a 5.7\% reduction in total loss over a harmonic-plus-noise baseline model, while providing interpretable parameters corresponding to physical phenomena. Complete code, model weights, and audio examples are openly available.
\end{abstract}

\begin{IEEEkeywords}
engine sound synthesis, 
differentiable signal processing, 
physics-informed neural networks, 
inductive biases, 
pulse-train synthesis, 
resonator modeling, 
neural audio synthesis
\end{IEEEkeywords}

\section{Introduction}

Engine sounds present an fundamental acoustic paradox: they exhibit distinctly harmonic spectral characteristics yet originate from inherently non-harmonic processes---discrete, explosive pressure pulses occurring at specific intervals. In a four-stroke engine, combustion events generate sharp pressure transients recurring at rates from 600 to over 8000~RPM (10--133~Hz). This creates acoustic phenomena with significant inharmonicity, extremely low fundamental frequencies down to 5~Hz, and rapid temporal sequences at intervals below 2~milliseconds. These properties demand synthesis approaches that can model both precision in timing and complexity in timbral evolution, beyond conventional musical audio assumptions.

Existing engine sound synthesis methods broadly follow two strategies: spectral modeling approaches that directly reconstruct observable acoustic characteristics through additive or sample-based synthesis~\cite{heitbrinkDesignDrivingSimulation2007, caoEngineOrderSound2020, jaglaSamplebasedEngineNoise2012, chenSynthesisingSoundCar2021, liRealTimeAutomotiveEngine2024}, and physics-based procedural methods that explicitly simulate combustion or mechanical processes~\cite{farnellDesigningSound2010, baldanPhysicallyInformedCar2015} but lack the adaptability and expressiveness of data-driven models.
Recent advances in neural audio synthesis, particularly Differentiable Digital Signal Processing (DDSP), have demonstrated remarkable capabilities in modeling complex audio phenomena through differentiable harmonic-plus-noise synthesis~\cite{engelDDSPDifferentiableDigital2020}. 
Adaptations for engine sounds have incorporated domain-specific features such as engine-phase conditioning and DCT-domain transient modeling \cite{lundbergDataDrivenProceduralAudio2020}, and enabled generation of static-RPM audio loops for sample-based synthesis applications 
\cite{lobatoMotor2SynthLeveragingDifferentiable2025}.

However, these DDSP-based methods also model the acoustic \textit{result}---the observed harmonic spectrum---rather than the physical \textit{cause}: the sequential pulse structure that generates this harmonicity through temporal periodicity.
This physical reality suggests that directly modeling the pulse structure may provide stronger inductive biases for neural synthesis architectures, yet existing methods do not implement it in a form amenable to gradient-based learning.

We propose the PTR model, which directly models the pulse train structure and exhaust system propagation underlying engine acoustics. The architecture integrates physics-informed inductive biases including harmonic decay, thermodynamic pitch modulation, valve-dynamics envelopes, and derived engine operating modes. Our differentiable implementation of the recursive Karplus-Strong algorithm enables gradient-based optimization of exhaust resonance simulation. 

By modeling the physical causes of engine sound rather than only their spectral manifestations, PTR provides a synthesis framework that improves reconstruction quality while yielding interpretable parameters corresponding to meaningful mechanical phenomena.

\section{Pulse-Train-Resonator Architecture}

\subsection{Overall Design Philosophy}

\begin{figure*}[htbp]
\centering
\includegraphics[width=\textwidth]{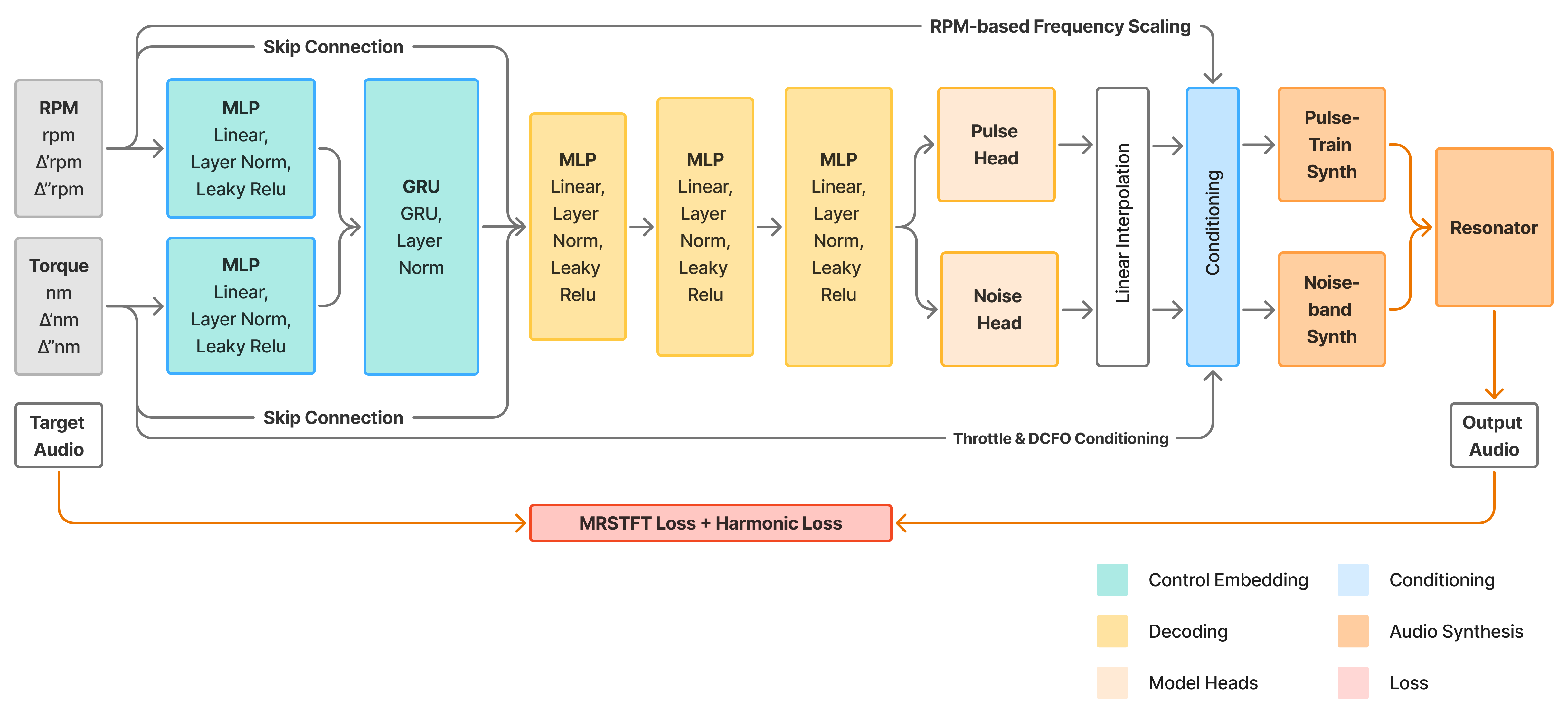}
\caption[Procedural Engines Model overview]{Control features (RPM, torque) and their deltas are temporally embedded via MLP blocks and GRU. Outputs are decoded into time-varying synth parameters through MLP blocks, converted to parameter ranges by specialized heads (Pulse, Noise), upsampled to audio rate and scaled by conditioning signals. Audio is synthesized via differentiable modules with network parameter updates calculated from multi-resolution spectral and harmonic losses.}
\label{fig:architecture_overview}
\end{figure*}

The PTR architecture (Figure~\ref{fig:architecture_overview}) implements a sequential processing pipeline that transforms engine control parameters (RPM, torque) into time-domain audio through three stages: (1) temporal control encoding with derived physical conditioning signals, (2) physics-informed pulse generation, and (3) exhaust resonance modeling. The architecture maintains full differentiability throughout, enabling end-to-end gradient-based optimization while embedding domain knowledge as architectural constraints.

\subsection{Input Feature Engineering}
In contrast to many musical sound sources, where pitch trajectories are largely direction-invariant, engine acoustics depend strongly on operational direction: identical RPM values yield distinct timbral characteristics during acceleration versus deceleration.

We augment control signals with temporal derivatives to capture dynamic behaviors:

\textbf{Engine speed deltas:} First-order difference $\Delta\text{RPM}(t) = \text{RPM}(t) - \text{RPM}(t-h)$ captures direction and rate of rotational speed changes, while second-order difference $\Delta^2\text{RPM}(t)$ identifies sudden dynamics from gear shifts or clutch engagement.

\textbf{Torque deltas:} First-order $\Delta\text{Nm}(t)$ distinguishes steady-state from transient load conditions, while second-order $\Delta^2\text{Nm}(t)$ captures abrupt mechanical events such as gear shifts and load reversals, manifesting as distinctive acoustic signatures.

Input features $\{\text{RPM}, \text{Nm}, \Delta\text{RPM}, \Delta\text{Nm}, \Delta^2\text{RPM}, \Delta^2\text{Nm}\}$ are averaged within frames at frame rate (125 Hz) and standardized using training-set statistics. 

\subsection{Inference and Conditioning Pathways}
All time-varying synthesis parameters are network-predicted; resonator coefficients are directly learned parameters, encouraging convergence on a single dataset-wide solution. Standardized features are encoded into frame-rate embeddings, decoded into synthesis parameters, and upsampled to audio rate (16~kHz) for waveform generation. In parallel, the original (non-standardized) RPM and torque trajectories are preserved at audio rate to derive physically meaningful conditioning signals applied directly before the synthesis stage.

\subsection{Physics-Informed Conditioning Signals}
\label{subsec:conditioning_signals}

Engine noise exhibits distinct spectral and temporal structures governed by operating regimes: during propulsion (positive torque), combustion generates rhythmic broadband transients and acoustic distortion from turbulent exhaust gas flow; during deceleration fuel cutoff (negative torque), combustion ceases while steady aeroacoustic noise persists as the engine is back-driven by the drivetrain.

Phase-agnostic STFT losses can capture RPM--torque--dependent spectral structure but cannot supervise cycle-synchronous modulations due to arbitrary frame alignment. Rather than learning these regimes implicitly from torque embeddings, we therefore explicitly encode them as architectural constraints. Torque polarity provides a direct physical indicator of operating mode, which we transform into gating signals:

\textbf{Throttle factor} activates combustion-related noise during propulsion:
\begin{equation}
g_{\text{thr}}(t) = \max(\text{torque}(t), \epsilon)^{0.7}
\end{equation}
where the sublinear exponent increases sensitivity at low torque, capturing acoustic response to initial throttle activation, with $\epsilon=0.02$ maintaining minimum gain during idle.

\textbf{DFCO factor} activates air flow noise during deceleration:
\begin{equation}
g_{\text{DFCO}}(t) = \max(-\text{torque}(t), \epsilon)
\end{equation}
These deterministic gating functions (applied in Section~\ref{sec:stochastic_augmentation}) enforce regime-specific activation of noise components and their temporal modulation behavior, providing explicit inductive bias that guides optimization toward physically plausible solutions.

\section{Differentiable Pulse Synthesis}

\subsection{Continuous Pulse-Train Derivation}

Pulse trains in their most simple form are sequences of discrete Dirac deltas, which can be approximated as continuous functions by Fourier series expansion using the sum of zero-phase cosines with fundamental frequency $T$: $\delta_T(t) \approx \frac{1}{T} + \frac{2}{T}\sum_{k=1}^K \cos(2\pi kt/T)$. However these analytical, unipolar signals misrepresent the physical reality that, rather than instantaneous steps, exhaust pulses exhibit rapid pressure \emph{gradients} that oscillate around equilibrium. We therefore employ a derivative-of-cosine representation that yields bipolar waveforms naturally capturing these pressure gradients:
\begin{equation}
\label{eq:cos_derivative}
\frac{d}{dt}\left[\sum_{k=1}^K a_k\cos(k\omega t)\right] = -\sum_{k=1}^K a_k k \omega \sin(k\omega t).
\end{equation}
We omit normalization by angular frequency $k \cdot \omega$ to avoid disproportionate gradient magnitudes for higher harmonics during backpropagation. 

\subsection{Physics-Motivated Pulse Shaping}

The base pulse formulation is further augmented with two physics-informed transformations:

\textbf{Pressure-release amplitude modulation} $E_i(\phi_i, t)$ models the rapid pressure release and subsequent decay during the exhaust event:
\begin{equation}
\label{eq:exponential_envelope}
E_i(\phi_i, t) = \left(1 - \exp(-\alpha_i(t)\phi_i)\right) \exp(-\beta_i(t)\phi_i),
\end{equation}
where $\phi_i \in [0,2\pi)$ is the phase in the firing-cycle for cylinder $i$, obtained by adding a fixed cylinder offset to the engine-cycle phase $\phi(t) = 2\pi \int f_0(t)\, dt$, with $f_0(t) = \mathrm{RPM}(t)/120$, and learnable coefficients $\alpha_i(t)$ and $\beta_i(t)$ control attack and decay rates modeling the asymmetric pressure transient produced when high-pressure gases are released.

\textbf{Thermodynamic phase modulation} captures frequency modulation from temperature-dependent sound propagation. Hot combustion gases ($\sim$800-1000°C) exhibit elevated sound speed ($c \propto \sqrt{T}$) compared to cooler residual gases in the manifold:
\begin{equation}
\tilde{\phi_{i}}(\phi_i, t) = 2\pi \cdot \left[\frac{\phi_i}{2\pi}\right]^{\nu_i(t)}, \quad \nu_i \in (0, 1].
\end{equation}
The exponential phase bending $\nu_i(t)$ compresses the pulse onset and stretches toward the end, reflecting that the leading edge of the pulse propagates faster than the trailing edge, creating a downward pitch trajectory as the pulse travels.

\subsection{Constrained Pulse Parameterization}

\begin{figure}[htbp]
\centering
\includegraphics[width=0.5\textwidth]{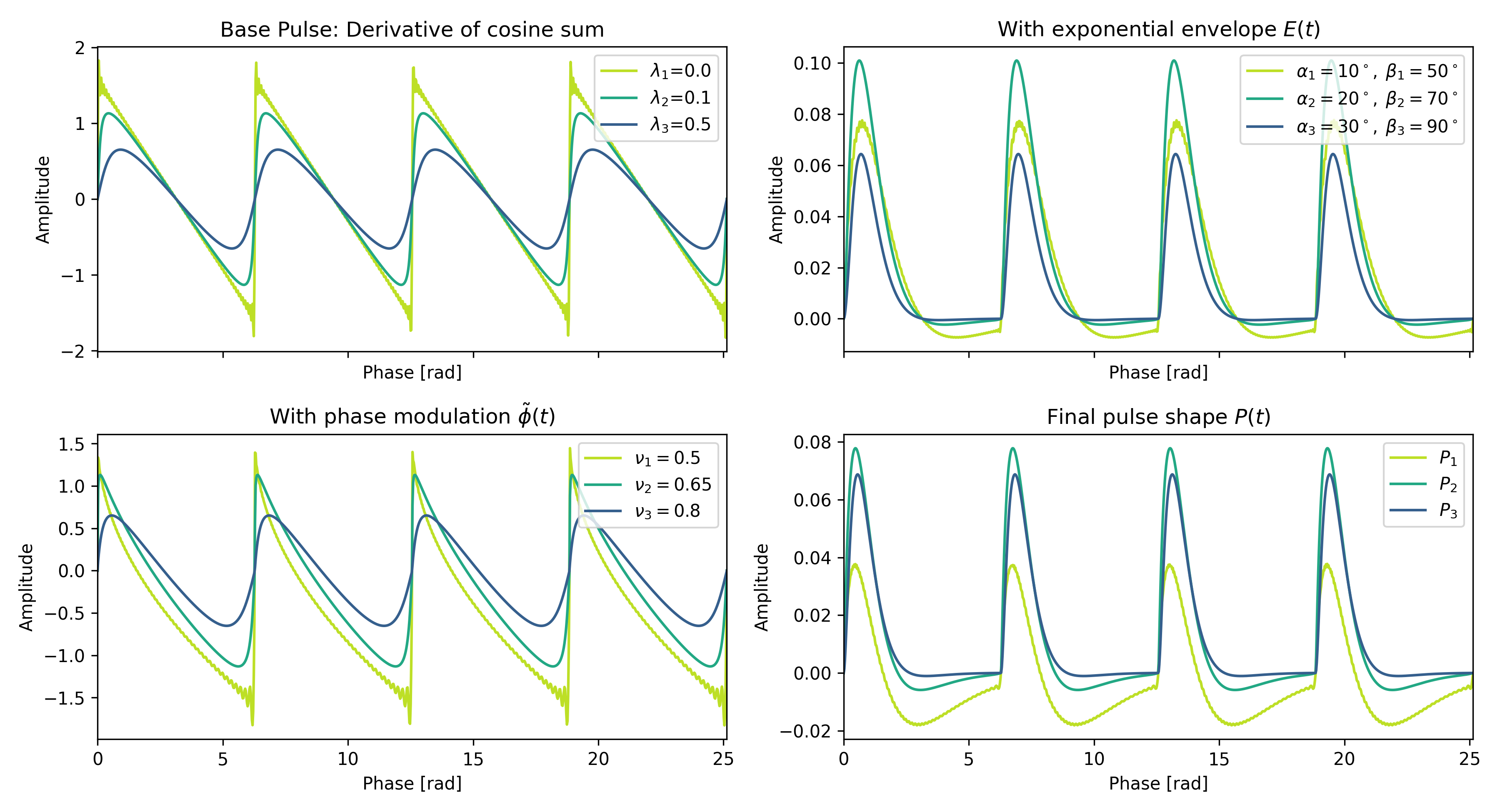}
\caption[Pulse shapes for different parameter settings]{%
Pulse shapes across varying parameter settings: (top-left) base pulses with varying harmonic decay $\lambda$; (top-right) exponential envelopes $E(t)$ with varying $\alpha, \beta$; (bottom-left) thermodynamic phase modulation $\tilde{\phi}$ with varying bending $\nu$; (bottom-right) final pulses $P(t)$ combining all elements. Color gradients distinguish corresponding parameter sets 1--3.
}
\label{fig:ptr_curves}
\end{figure}

The complete physically-informed pulse train $P_i$ combines these elements:
\begin{equation}
\label{eq:combined_pulse_train}
P_i(\phi_i, t) = E_i(\phi_i, t) \cdot c_i(t) \cdot \left[\sum_{k=1}^K a_{i,k}(t) \sin(k \tilde{\phi}_i(\phi_i,t))\right],
\end{equation}
where $c_i(t)$ are per-cylinder gains, and $a_{i,k}(t)$ are a normalized amplitude distribution incorporating an exponential harmonic decay ($a_{i,k}(t) \propto e^{-0.5k \lambda_i(t)}$). While the amplitudes directly parameterize the in-cycle pulse shape, the exponential decay constrains them to always realize a bipolar localized pulse (conjugate Poisson kernel family), guaranteeing impulsivity by construction. Unlike standard harmonic synthesizers, where independent amplitudes shape a \emph{timbral envelope}, our coefficients directly parameterize a \emph{pulse shape} in time. Figure~\ref{fig:ptr_curves} illustrates such pulse shapes across varying parameter settings. 

\subsection{Stochastic Augmentation}
\label{sec:stochastic_augmentation}

The deterministic pulse $P_i$ is augmented with three noise sources that add plausible stochasticity to idealized pressure waves while providing broadband excitation for the resonator model (Section~\ref{sec:resonator}): 
(1) \textbf{turbulence in exhaust gas flow} distorts the pulse signal through stochastic amplitude modulation, (2) \textbf{intake system pulsations} capture additional impulsive events from valve dynamics and air column reversals, and (3) \textbf{steady air flow} represents aeroacoustic radiation during DFCO (engine operating as air pump):
\begin{equation}
\begin{aligned}
\label{eq:pulse_train_noise}
\hat{P}_i(\phi_i, t) &= P_i(\phi_i, t) \cdot \Bigl[1 + \alpha_{\text{turb}} \cdot g_{\text{thr}}(t) \cdot \eta(t)\Bigr] \\
&\quad + \eta(t) \cdot \Bigl[E(\phi, t) \cdot g_{\text{thr}}(t) + g_{\text{DFCO}}(t)\Bigr]
\end{aligned}
\end{equation}
where $\eta(t) = \sum_{b=1}^B \alpha_b(t) \eta_b(t)$ denotes an ERB-spaced cosine-filtered noise bank weighted by learnable time-varying gains $\alpha_b(t)$, $\alpha_{\text{turb}}$ controls turbulence distortion depth, $E(\phi,t)$ is an exponential envelope of the form (Eq.~\ref{eq:exponential_envelope}) synchronized to the global engine cycle phase $\phi$, and $g_{\text{thr}}(t)$, $g_{\text{DFCO}}(t)$ apply the conditioning from Section~\ref{subsec:conditioning_signals}.

\subsection{Multi-Cylinder Synthesis}

Each cylinder firing is modeled by an independent periodic impulse train $\hat{P}_i(\phi_i,t)$ (Eq.~\ref{eq:pulse_train_noise}), with phase offsets following a standard firing order $[1,5,4,8,6,3,7,2]$ of a V8 petrol engine configuration (two banks of four cylinders in a V geometry) and refined by learned per-cylinder timing adjustments bounded to $\pm 40^\circ$ crank angle. Cylinder outputs are summed within banks (cylinders 1--4 on the left, 5--8 on the right) before resonator processing.

\section{Differentiable Exhaust Resonance}
\label{sec:resonator}

\subsection{Karplus-Strong Algorithm Adaptation}
Exhaust system acoustics involve wave reflections, comb filtering, and complex interactions between oncoming pulses and pressure wave propagation in tubes of varying geometry. We model these phenomena through Karplus-Strong feedback delay lines \cite{karplusDigitalSynthesisPluckedString1983}, expressed as:
\begin{equation}
\label{eq:ks_difference}
y[n] = x[n] + g \cdot h[n-L],
\end{equation}
where $L \in \mathbb{N}$ is the delay length in samples (corresponding to fundamental resonance frequency $f_0 \approx f_s / L$), $g$ is the feedback gain, and $h[n-L]$ is a filtered delayed signal:
\begin{equation}
h[n-L] = \alpha \cdot y[n-L] + \beta \cdot y[n-L-1],
\end{equation}
This two-coefficient formulation ($\alpha, \beta$) provides greater flexibility than traditional Karplus-Strong ($\alpha = \beta = 0.5$), enabling independent control over filter characteristics within the feedback path.

\subsection{Enabling Gradient-Based Optimization}

Direct implementation of recursive filters creates backpropagation-through-time challenges: long sequential dependency chains cause vanishing gradients and prevent parallel computation. We address this by recognizing that Karplus-Strong delay-line feedback is a constrained instance of the general all-pole filter
\begin{equation}
y[n] = x[n] - \sum_{i=1}^{M} a_i[n] \cdot y[n-i],
\end{equation}
which reduces to the Karplus-Strong form when 
\begin{align}
a_L &= -g \cdot \alpha, \quad a_{L+1} = -g \cdot \beta \\
a_i &= 0 \quad \text{for } i \notin \{L, L+1\}.
\label{eq:kp_spars_coeffs}
\end{align}
This imposes two structural constraints: \textit{sparsity} (only 2 of $M$ coefficients non-zero) and \textit{positioning} (non-zero coefficients at delays $L$ and $L+1$).
Following Yu et al. \cite{yuDifferentiableAllpoleFilters2024}, we reformulate the forward pass as a non-recursive infinite impulse response, eliminating sequential dependencies while maintaining identical filter response. This enables efficient gradient computation without unrolling recursive dependencies.

\subsection{Implementation Details}

\subsubsection{Differentiable delay selection}
Delay $L$ is selected via Gumbel-Softmax with a straight-through estimator (hard one-hot forward pass). The two active coefficients $(a_1, a_2)$ are placed at the selected delay $L$ (and $L+1$), so the coefficient vector is sparse by construction (Eq.~\ref{eq:kp_spars_coeffs}).

\subsubsection{Stability guarantees} Filter stability is ensured through reflection coefficient parameterization. The network predicts unconstrained parameters converted to reflection coefficients $k_i = \tanh(\theta_i)$, constraining $|k_i| < 1$ which theoretically guarantees all poles lie within the unit circle. Direct-form coefficients are derived through established conversions \cite{smithIntroductionDigitalFilters2007}:
\begin{equation}
a_1 = k_1(1 - k_2), \quad a_2 = k_2.
\end{equation}
Additional numerical stability during GPU training is achieved by clamping $|a_2| < 0.999$ and constraining $a_1$ within the stability triangle bounds. Feedback gain is integrated into coefficients: $\alpha_{\text{eff}} = a_1 \cdot \sigma(g)^{0.35}$, $\beta_{\text{eff}} = a_2 \cdot \sigma(g)^{0.35}$, where the sublinear exponent biases the model toward active resonator utilization rather than bypass through zero gain.

\subsubsection{Minimum delay enforcement}
Left-padding with $L_{\min}$ zeros offsets the selected delay by $L_{\min}$, preventing short-delay, non-resonant filtering and ensuring physically plausible fundamental frequencies.

\subsubsection{Resonator configuration} Two independent resonators process cylinder bank outputs (modeling distinct manifold paths), with processed signals combined in a final shared resonator (common exhaust pipe). Their fixed (non-time-varying) parameterization reflects the geometric stability of the exhaust system.

\section{Training and Evaluation}
\label{sec:training}

\subsection{Dataset and Training Setup}

We train PTR on three subsets (A, B, C) of the Procedural Engine Sounds Dataset~\cite{doerflerAnalysisDrivenProceduralGeneration2026}, each containing approximately 2.5 hours of audio. The subsets represent different petrol engine acoustics: (A) inline-four cylinder configuration with predominantly harmonic components, (B) V8 with moderate stochastic perturbations and low-frequency resonant exhaust, and (C) V8 with pronounced mid-range frequencies, strong harmonic deviations and metallic resonances transforming broadband exhaust noise components. This progression ($A < B < C$) increases spectral and temporal complexity, enabling evaluation of generalization capabilities.
Each subset undergoes a 90/10 train-validation split.

Audio is processed in batches of 8 containing 65,536-sample mono chunks (4 seconds at 16 kHz) extracted with 50\% overlap. Control signals are downsampled to 125 Hz model frame rate and standardized using training-set statistics. Additionally, non-standardized RPM and torque are preserved at audio rate for conditioning signal derivation. Training employs AdamW optimizer (learning rate $1 \times 10^{-3}$, weight decay $1 \times 10^{-2}$) with one-cycle scheduling over 100 epochs ($\approx$45,000 steps). 

\subsection{Loss Function Design}
We employ multi-resolution STFT loss with FFT sizes ranging from $32{,}768$ down to $32$ samples (75\% overlap, Hann window), capturing spectral structure across time-frequency resolutions. The loss combines spectral convergence, linear magnitude, log-magnitude, and spectral energy terms with equal weighting and scale-invariant normalization ensuring equal contribution across all resolutions.

An additional harmonic loss supervises frame-wise energy along predicted engine-order harmonics, inspired by Campbell diagrams from rotating machinery analysis. Energies are computed from magnitude spectrograms masked along harmonic tracks derived from instantaneous RPM, using high spectral resolution (FFT $65{,}536$, window $16{,}384$, hop $256$) to minimize spectral leakage and isolate harmonic regions at low fundamental frequencies.

\subsection{Quantitative Results}

Table~\ref{tab:best_val_losses} presents validation performance comparing PTR against a Harmonic-Plus-Noise (HPN) baseline~\cite{serraSpectralModelingSynthesis1990}, adapted for engine sound synthesis using the identical encoder-decoder architecture as PTR but replacing pulse-train and resonator synthesis with sinusoidal-plus-filtered-noise synthesis~\cite{doerflerNeuralEngineSound2025}. PTR consistently outperforms the baseline across all three datasets, with improvements ranging from 3.8\% to 7.6\% in total validation loss. Mean performance shows 5.7\% total loss reduction and 21\% improvement in harmonic reconstruction, indicating that physics-informed pulse constraints promote more generalizable representations.
\begin{table}[ht]
\centering
\caption[Validation Loss Comparison]{Validation Loss Comparison}
\begin{tabular}{lccc|ccc}
\toprule
 & \multicolumn{3}{c}{HPN} & \multicolumn{3}{c}{PTR} \\
Dataset & Harmonic & STFT & Total & Harmonic & STFT & Total \\
\midrule
A & 0.107 & 1.781 & 0.944 & \textbf{0.090} & \textbf{1.649} & \textbf{0.872} \\
B & 0.059 & 1.824 & 0.943 & \textbf{0.055} & \textbf{1.754} & \textbf{0.907} \\
C & 0.166 & 2.093 & 1.132 & \textbf{0.117} & \textbf{2.017} & \textbf{1.069} \\
mean & 0.111 & 1.899 & 1.006 & 0.088 & 1.807 & 0.949 \\
\bottomrule
\end{tabular}
\label{tab:best_val_losses}
\end{table}

PTR's superior harmonic reconstruction despite not modeling harmonics directly suggests that cycle-locked, decay-constrained pulse parameterization acts as a stronger inductive bias for impulsive periodic sources than free per-harmonic amplitude modeling.
Both models demonstrate rapid early convergence within the first 10,000 training steps. PTR maintains consistent performance across three diverse engine configurations (Section~\ref{sec:training}), with successful generalization to Dataset A despite the model's V8 firing-order architectural prior demonstrating robustness to configuration mismatches.

\subsection{Qualitative Results}
Beyond quantitative metrics, informal listening reveals how PTR's physics-informed structure manifests in synthesis quality\footnote{Audio examples: \url{https://rdoerfler.github.io/ptr-model-page/}}. The model produces authentic engine character: RPM-dependent harmonicity, load-dependent noise coupling, complex tonal evolution during acceleration and gear shifts, and distinct acoustic signatures for throttle operation (sharp rhythmical noise bursts) versus deceleration fuel cutoff (steady turbulent flow).

Several behaviors emerge that were not explicitly designed. During clutch disengagement, combustion events become intermittent and resume synchronization upon re-engagement---a mechanical transition arising from the pulse-train architecture rather than from any dedicated training signal. The cycle-locked pulse structure also yields a natural articulation gradient: individual combustion events are clearly audible at low RPM and blend into dense harmonic textures at high RPM. The Karplus-Strong resonators produce convincing exhaust-pipe resonances, though their spectral character occasionally deviates from the target.

\section{Conclusion}
We show that physics-informed inductive biases and direct modeling of pulse shapes -- rather than spectral targets -- offer an effective approach to neural engine sound synthesis. By integrating domain knowledge at multiple architectural levels (parameterized pressure pulse generation, firing-order sequencing, and differentiable Karplus-Strong resonators) the PTR architecture improves both spectral reconstruction and harmonic accuracy. Beyond these metrics, it exposes interpretable parameters that map directly to physical phenomena such as valve timing, phase modulation, and exhaust resonance, offering insight into how mechanical properties shape timbre. Future directions include validation on real-world recordings, audio-driven control prediction for end-to-end training on unannotated collections, improved resonator timbre matching, and extension to broader vehicle acoustics such as backfiring, turbo noise, and drivetrain components.

\bibliographystyle{IEEEtran}
\bibliography{refs}

\end{document}